\documentclass[aps,amsmath,amssymb,superscriptaddress,longbibliography,notitlepage]{revtex4-2}

\usepackage{amsmath}
\usepackage{mathscinet}
\usepackage{amssymb}
\usepackage{fancyhdr}
\usepackage{emptypage} 
\usepackage{float}
\usepackage{graphicx}
\usepackage[utf8]{inputenc}
\usepackage{indentfirst}
\usepackage{hyperref} 
\usepackage[]{xcolor}
\newcommand{\PhiV}{\Phi^\text{\tiny{V}}}
\newcommand{\PhiK}{\Phi^\text{\tiny{K}}}
\newcommand{\GV}{G^\text{\tiny{V}}}
\newcommand{\GK}{G^\text{\tiny{K}}}
\newcommand{\FV}{F^\text{\tiny{V}}}
\newcommand{\FK}{F^\text{\tiny{K}}}
\newcommand{\xst}{x_\text{\tiny{st}}}
\newcommand{\dxst}{\dot{x}}

\begin{document}

\title{Aging in some opinion formation models: a comparative study}
\author{Jaume Llabres}
\affiliation{
Instituto de Física Interdisciplinar y Sistemas Complejos IFISC (CSIC-UIB), Campus Universitat de les Illes Balears, 07122 Palma de Mallorca, Spain}
\author{Sara Oliver-Bonafoux}
\affiliation{
Instituto de Física Interdisciplinar y Sistemas Complejos IFISC (CSIC-UIB), Campus Universitat de les Illes Balears, 07122 Palma de Mallorca, Spain}
\author{Celia Anteneodo}
\affiliation{Department of Physics, Pontifical Catholic University of Rio de Janeiro, PUC-Rio,
Rua Marquês de São Vicente, 225, 22451-900 Rio de Janeiro, Brazil}
\affiliation{
National Institute of Science and Technology for Complex Systems, INCT-SC, Rio de Janeiro, Brazil}
\author{Ra\'ul Toral}
\affiliation{
Instituto de Física Interdisciplinar y Sistemas Complejos IFISC (CSIC-UIB), Campus Universitat de les Illes Balears, 07122 Palma de Mallorca, Spain}

\begin{abstract}
Changes of mind can become less likely the longer an agent has adopted a given opinion state. This resilience or inertia to change has been called ``aging''.
We perform a comparative study of the effects of aging on the critical behavior of two standard opinion models with pairwise interactions.
One of them is the voter model, which is a two-state model with a dynamic that proceeds by simple social contagion, 
another is the so-called kinetic exchange model, which allows a third (neutral) state and the formed opinion depends on the previous opinions of both interacting agents. 
Furthermore, in the noisy version of both models, random opinion changes are also allowed, regardless of the interactions. 
Due to aging, the probability of change diminishes with the age, and to take this into account we consider algebraic and exponential kernels.
We investigate the situation where aging acts only on pairwise interactions. 
Analytical predictions for the critical curves of the order parameters are obtained for the opinion dynamics on a complete graph, 
in good agreement with agent-based simulations. 
For both models considered, consensus is optimized by an intermediate value of the parameter that rules the rate of decrease of the aging factor.
\end{abstract}
\keywords{Aging, opinion formation model, voter model, kinetic-exchange model, social dynamics} 
\maketitle


\section{Introduction}

As well known, the ideas and concepts coming from the Statistical and Nonlinear Physics toolbox have been successful to understand social phenomena~\cite{Castellano:2009,SanMiguel2020b}. The main goal is to explain a ``macroscopic'', emergent, behavior in terms of the ``microscopic'', individual's, components in the same way that one derives the laws of macroscopic physical systems, e.g. the equation of state, from microscopic information such as the forces amongst particles. One lesson we have learnt from the modern development of Statistical Physics is that the emergent behavior, manifested through macroscopic phase transitions, is largely independent on some details of the microscopic interactions, and that there are some universal features that only depend on a few ingredients, such as dimensionality, symmetries, etc. It is in this spirit that the modeling of social systems has relied on the development of very simple agent-based models that take as starting points some stylized facts about the interactions amongst agents. Consider, for example, the subject of the evolution of cultures. %
In a seminal paper, Axelrod asked the question ``If people tend to become more alike in their beliefs, attitudes, and behavior when they interact, why do not all such differences eventually disappear?''~\cite{Axelrod1997}. He then introduced a model whose basic assumption is that interactions amongst individuals lead to an augmentation in the similarity between them. A detailed analysis~\cite{Castellano:2000} shows that depending on the initial diversity there is a phase transition between an ordered monocultural state and a disordered multicultural state, demonstrating the apparent paradox of the emergence of global polarization through the mechanism of local convergence. It was then established that the dimensionality of the space of interactions was a relevant variable of the model~\cite{Klemm2003}. Similar ideas appear in other areas. In the context of opinion formation, for example, one expects, on the one hand, that social influences reduce differences between individuals by social contagion but, on the other, interactions can also produce differentiation by repulsive forces, associated to anticonformist, "contrarian", agents that tend to deviate from the behavior adopted by the neighbors in the network of connections~\cite{Galam2004}. Another example is Galam's model for minority opinion spread~\cite{Galam2002} that, although based on a local majority rule, can lead to an advantage of the hostile minority views. 

Diverse other mechanisms can compete with the tendency to uniformity promoted by social contagion. One is the 'noisy' behavior, ruled by a 'social temperature' parameter, associated to imperfect imitations due to random factors or independent choices, such that random changes of opinion interpreted as idiosyncratic behavior can occur. Standard models that study the effect of independence are the noisy voter model~\cite{carro2016noisy,peralta2018stochastic,khalil2018zealots,Kirman1993} and the noisy kinetic-exchange model~\cite{Crokidakis2014,Vieira2016}. In the former, agents adopt opinions $\pm 1$, copying the opinion of a random contact, in the latter, the states $\pm 1$ and 0 can be adopted by a rule that takes into account the current opinion of both interacting contacts. In both cases the rule of social influence acts with probability $1-a$, while otherwise, with probability $a$, a random change can occur. That is, these models rely on the assumption that opinion dynamics is mainly governed by mechanisms of imitation or social contagion that mold opinion formation through the interaction between individuals, but random factors are always present. In both cases, noise opposes the system ability to reach a self-organized state with a winning opinion. Furthermore, both present a critical value $a_\mathrm{c}$ below which order can be achieved, a main difference being that $a_\mathrm{c} \to 0$ in the thermodynamic limit, when the number of agents tends to infinity, for the noisy voter, while $a_\mathrm{c}$ remains finite in the kinetic-exchange model in the same limit. 

Another ingredient that has been proven to be relevant in the modeling of social systems is that of ``aging'', or the larger resistance an agent offers to change its state of opinion the longer it has been holding the current state. Initially termed as ``inertia'' ~\cite{Stark2008}, it was shown that the slowing down of the microscopic dynamics induced by aging actually decreased the time needed to reach the macroscopically ordered state state of consensus. In this respect, and aligned with previous work~\cite{Stark2008,fernandezgracia2011,perez2016,Artime2018,Abella2022,Abella2023}, we will consider that aging acts on the mechanism associated to social contagion. 

It is the aim of this paper to compare the effects of aging on the two above-mentioned paradigmatic representatives of the class of noisy opinion models, the voter model and the kinetic-exchange model, highlighting the similarities and differences amongst them. To this end we will consider an all-to-all interaction scheme in which every agent is connected to every other agent, and will develop an adiabatic approximation able to provide us with the phase diagram, including the location of the critical points separating regions of consensus with disordered regions. Afterwards, we compare the location of the critical points for both models. 

The rest of the paper is organized as follows: In Section~\ref{sec:model}, we introduce a general setup to study models of opinion dynamics under the presence of idiosyncratic, random, changes of opinion, as well as the interaction between the agents modulated by an aging mechanism. In Section~\ref{sec:noisymodel} we derive a mean-field type approach for the noisy voter model under an adiabatic approximation, leaving the more technical details for Appendix~\ref{sec:app:nvm}. In the same section we also compare with the results of numerical simulations for two different functional forms for the aging probability, algebraic and exponential, focusing mainly on the magnetization, measuring the amount of order, or amount of consensus, in the system. In Section~\ref{sec:kinetic}, after reviewing briefly the main results for the noisy kinetic-exchange model, we compare the results for the magnetization and the dependence of the critical value of the noise intensity as a function of a parameter measuring the rate of decrease of the aging probability, for the voter and the kinetic-exchange model. Finally, in Section~\ref{sec:final} we end with some conclusions and outlook for future work. 

\section{Models of noisy opinion dynamics with aging}
\label{sec:model}
 
In this work we analyze two models that have been widely used as prototypical examples in the study of the dynamics of opinion formation in social systems: the noisy voter model and the noisy kinetic-exchange model. Both models consider a set of $N$ agents endowed with an internal state variable $s_i,\,i=1,\cdots,N$, representing the possible positions of agent $i$ concerning a given topic. For the voter model the internal variable can take two possible values $s_i\in\{-1,+1\}$, representing a position against or in favor of the topic. The kinetic-exchange model introduces a third neutral state such that $s_i\in\{-1,0,+1\}$. This internal variable evolves due to two different generic mechanisms that act stochastically: idiosyncratic changes and social influence. The difference between them is that the former changes randomly the state variable $s_i$ independently on the states of other agents, while the latter changes $s_i$ following a rule that depends on the state variables of another agent $s_j$. In the voter model the social rule is that $s_i \to S(s_i,s_j)=s_j$, modeling the mechanism of imitation. In the kinetic-exchange model the social rule is $s_i \to S(s_i,s_j)= {\rm sgn}[s_i+s_j]$, where ${\rm sgn}[s]$ is the sign function. The rule is such that neutral agents can not modify the opinion of another agent, and agents with a well defined opinion (either positive or negative) can convince a neutral agent or turn into the neutral state an agent with the opposite opinion. 

Previously, both models have been considered under the influence of aging in the social mechanism~\cite{peralta2018stochastic, OurArXiv2023}. It is the goal of this paper to generalize some of the previous work and to compare the effects that aging induces in both models. 

To be precise, let us spell out in detail the evolution rules for these models. Initially we assign a random value to each state variable $s_i$ and set all internal times $\tau_i$ to zero. Then,
\begin{enumerate}
\item At each iteration, an agent $i$ is randomly selected.
\item With probability $(1-a)$ the social rule is chosen: With probability $q(\tau_i)$, modeling the persistence or reaction of the individual $i$ to change as a function of its age $\tau_i$, a neighbor $j$ is randomly selected and the opinion of agent $i$ is modified according to the rule $s_i \to S(s_i,s_j)$. 

\item Otherwise, with probability $a$, the idiosyncratic rule is chosen: the state $s_i$ is replaced by a randomly selected value among all possible opinion states. 

\item Irrespective of the update mechanism actually used by agent $i$ (random change or pairwise interaction, with aging or not) its age $\tau_i$ is updated in the following way: If the state $s_i$ has changed, then age is reset, i.e., $\tau_i\to 0$, otherwise, age is incremented in one unit, i.e., $\tau_i\to\tau_i+1$. 
\end{enumerate}

We will restrict ourselves in this work to the all-to-all connectivity, where each agent is linked to every other agent. Hence, the set of neighbors of an agent $i$ is the whole set of agents (excluding itself). Time is measured in Monte Carlo steps, such that one unit of time corresponds to $N$ agent selections for updating.

Although other forms are possible, in the present work, we adopt the following two general functional forms for the probabilities (we use $q(\tau)\equiv q_\tau$, for brevity in the notation), 
\begin{eqnarray}\label{eq:qtaupot}
&\text{algebraic decay}:\hspace{20pt}&q_\tau=\frac{q_\infty\tau+q_0\tau^*}{\tau+\tau^*},\\
\label{eq:qtexp}
&\text{exponential decay}:\hspace{20pt}&q_\tau=q_\infty+(q_0-q_\infty)e^{-\tau/\tau^*},
\end{eqnarray}
where $q_0$ and $q_\infty$, satisfying
$0\le q_\infty < q_0\le 1$, denote, respectively, the initial ($\tau=0$) and asymptotic ($\tau\to\infty$) values of $q_\tau$, and $\tau^*$ is a parameter dictating the rate of decrease of the aging-probability, i.e. a larger value of $\tau^*$ implies a slower decay. Both forms have been considered before in the context of the study of the way that the voter model (without idiosyncratic changes) approaches the asymptotic consensus state~\cite{Peralta:2020,Baron:2022}. A previous study of the noisy voter model~\cite{Artime2018} was limited to the algebraic case with $q_0=1/2,\,\tau^*=2$. The aging-less case is recovered formally taking the limit $\tau^*\to\infty$, implying $q_\tau=q_0$ for all values of the age $\tau$. Our theoretical treatment is rather general, but we adopt the values $q_0 = 1$ and $q_\infty=0$ in the numerical simulations.

\section{Noisy voter model with aging}\label{sec:noisymodel}
In this section we present the results obtained for the noisy voter model with aging. We build over the treatment of Ref.~\cite{Artime2018} but present a more general one, valid for arbitrary forms of the aging-update probability. 

In this model each agent $i$ is characterised by its binary state variable, $s_i \in \{-1, +1\}$, and its age or residence time in state $s_i$, $\tau_i$. We denote by $x^\pm_\tau(t)$ the fractions of agents in states~$\pm1$ and with age $\tau$, such that $x(t) = \sum_{\tau = 0}^\infty x^+_\tau(t)$ is the total fraction of agents in state $+1$ at time $t$, and $1-x(t)=\sum_{\tau = 0}^\infty x^-_\tau(t)$ is the total fraction of agents in state $-1$ at time $t$. As detailed in Appendix~\ref{sec:app:nvm}, it is possible to derive a closed evolution equation for $x(t)$. The procedure starts by writing down rate equations for the densities $x^\pm_\tau(t)$. One then performs an adiabatic approximation that assumes that the evolution of $x^\pm_\tau(t)$ is slaved to that of $x(t)$ which evolves at a longer time scale. Under this approximation, the equation for the dynamical evolution of the fraction $x(t)$ of agents in state $+1$ at time $t$ is
\begin{equation}\label{eq:dxdtVP}
\begin{split}
 \frac{dx}{dt} &= \GV(x),\\
 \GV(x)&=\frac{a}{2}(1-2x) + (1-a) x (1-x) [\PhiV(x) - \PhiV(1-x)],
 \end{split}
\end{equation}
where the function $\PhiV(x)$ is defined as
\begin{equation} 
\PhiV(x) = \frac{\sum_{\tau=0}^\infty q_\tau \FV_\tau(x)}{\sum_{\tau=0}^\infty \FV_\tau(x)},
\end{equation}
with 
\begin{equation} \label{eq:FVP}
\FV_0(x)=1,\quad
\FV_\tau(x) = \prod_{k=0}^{\tau-1}\gamma_2(q_k\, x,a),\quad \tau\ge 1,
\end{equation}
and
\begin{equation} 
\label{eq:Xi_def}
\gamma_n(z,a)= \frac{a}{n}+(1-a)(1-z).
\end{equation}
The steady state-solutions $\xst$ are obtained by setting the time derivative of the density $x(t)$ equal to zero, or $\GV(\xst)=0$, according to Eq.~\eqref{eq:dxdtVP}. Note that due to its very nature, the adiabatic approximation does provide the correct values of these steady-state solutions. This is because the exact calculation of the steady-state solutions requires that the rates of change of all densities are set to zero. This leads to the condition $\GV(\xst)=0$ which is also a result of the adiabatic approximation. What the approximate dynamical equation \eqref{eq:dxdtVP} provides is the stability of the different steady-state solutions as determined by the sign of the derivative $\frac{d\GV(x)}{dx}$ evaluated at $\xst$.

In the aging-less case, $q_\tau=q_0,\,\forall \tau$, it is $\PhiV(x)=q_0$ and the only fixed point of Eq.~\eqref{eq:dxdtVP} is $\xst=1/2$, which is stable. This indicates that the disordered phase is the only asymptotic solution. It is clear from the structure of Eq.~\eqref{eq:dxdtVP} that $x_\text{st}=1/2$ is, trivially, also a steady-state solution for the aging situation, for an arbitrary function $\PhiV(x)$. The stability of this trivial solution and the existence of other steady-state solutions depend solely on the function~$\GV(x)$. 

For the algebraic functional form of the aging-update probabilities given by Eq.~\eqref{eq:qtaupot}, it is possible, as explained in Appendix~\ref{sec:app:Ftau}, to obtain an analytical form for the function $\PhiV(x)$ in terms of hypergeometric functions. However, even in this case, the non-trivial steady state values and their stability have to be obtained numerically. For the exponential form of the aging-update probabilities given by Eq.~\eqref{eq:qtexp}, it does not seem to be possible to express $\PhiV(x)$ in terms of other known functions, but we outline in Appendix~\ref{sec:app:Ftau_exp} a convenient numerical algorithm for its evaluation.

Both for the algebraic and the exponential forms for the aging-update probabilities, the numerical analysis concludes that, in the case $q_0=1,\,q_\infty=0$, the solution $x_\text{\tiny st}=1/2$ is stable for $a>a_\mathrm{c}(\tau^*)$ and unstable for $a<a_\mathrm{c}(\tau^*)$ and that a new pair $\xst,1-\xst$ of symmetric stable solutions emerge for $a\le a_\mathrm{c}(\tau^*)$. These two solutions share the same stability as it can be easily proved that $\left.\frac{d\GV(x)}{dx}\right|_{x=\xst}=\left.\frac{d\GV(x)}{dx}\right|_{x=1-\xst}$. This change of stability of the symmetric solution $x_\text{\tiny st}=1/2$ is understood as a phase transition between a disordered phase at $a>a_\mathrm{c}(\tau^*)$ where both opinions $\pm1$ coexist in equal proportion, to an ordered, consensus, phase for $a<a_\mathrm{c}(\tau^*)$ where one of the opinions is majoritarian. The existence of this phase transition is one of the main results presented in Ref.~\cite{Artime2018} for a particular form of the aging probability (corresponding to the algebraic case with $q_0=1/2,\,\tau^*=2$). The same transition was understood as a mapping of the problem with aging to another one with suitable nonlinear rates in the social mechanism~\cite{Artime2019}. We now present a general study for both the algebraic and exponential dependence of the aging probability and different values of the parameter $\tau^*$.

\begin{figure}[h!]
 \includegraphics[width=0.45\textwidth]{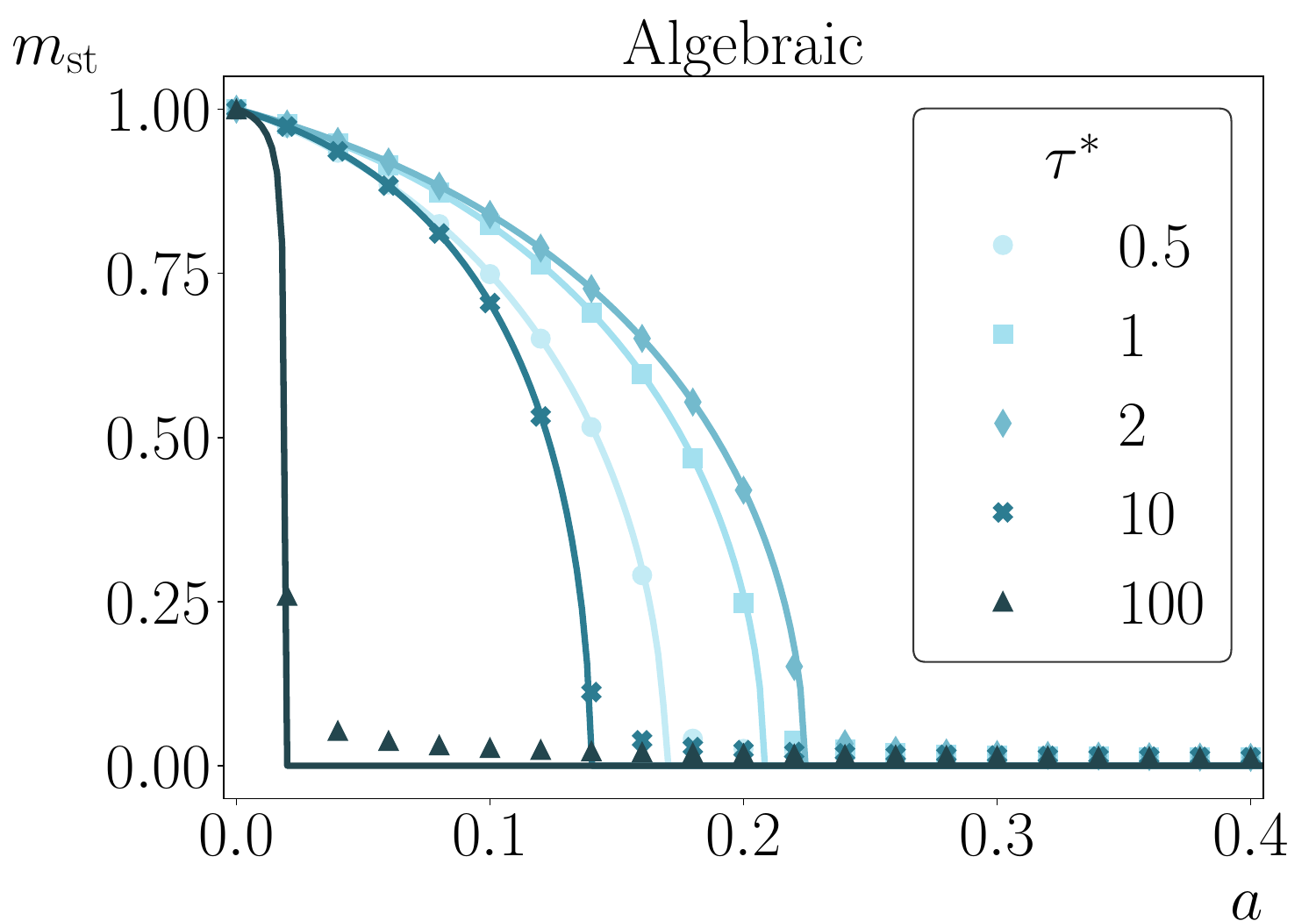}
 \includegraphics[width=0.45\textwidth]{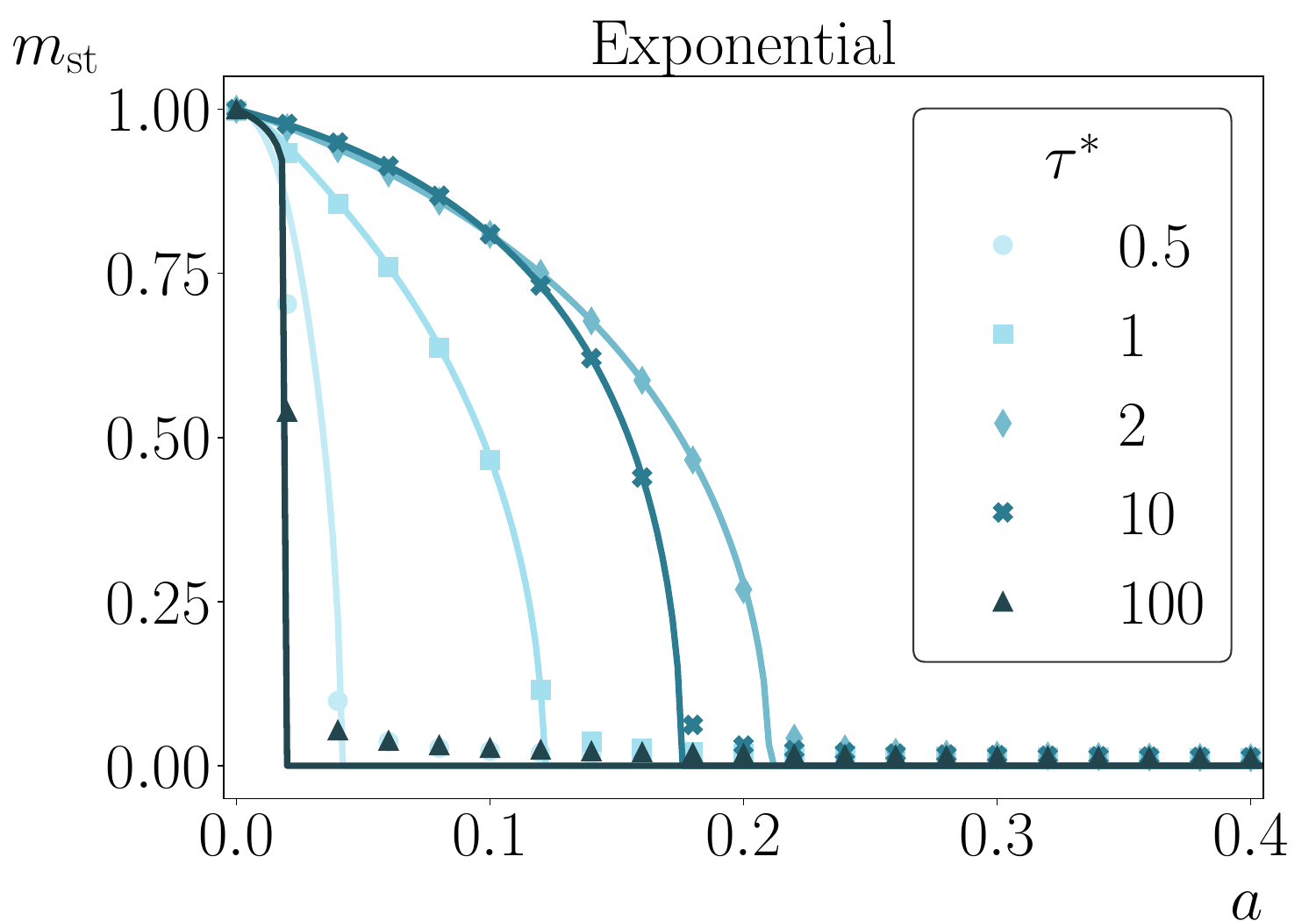}
\caption{Noisy voter model with aging. Magnetization, $m_\mathrm{st} = |2x_\mathrm{st} - 1|$, as a function of the noise intensity $a$ for different values of the $\tau^*$ parameter, for the algebraic and exponential kernels. Solid lines correspond to the theoretical results, while symbols come from numerical simulations with $N = 10^4$ agents, averaged over $5 \times 10^6$ MCS after a transient of $5 \times 10^6$ MCS. Notice the non-monotonic dependency of the critical point with the parameter $\tau^*$.}\label{fig:NVMP_mstvsa}
\end{figure}

In Fig.~\ref{fig:NVMP_mstvsa} we plot the {\slshape magnetization}, $m_\text{\tiny{st}}=|2\xst-1|$, as a function of the noise intensity,~$a$, both for the algebraic and the exponential functional forms of the aging update probability, and for several values of the parameter $\tau^*$. In the same graph we plot the results of numerical simulations of the stochastic rules for the agent-based model detailed in Section~\ref{sec:model}. The simulations agree remarkably well with the analytical results, validating the adiabatic approximation introduced in the analysis. There are some deviations with respect to the analytical calculation for values of the probability parameter $a$ which are close to the critical value $a_\mathrm{c}$. This is attributed to the necessary consideration of a finite number of agents $N=10^4$ in the numerical simulations, while the theoretical treatment considers the thermodynamic limit $N\to\infty$.

For a given value of $\tau^*$, the critical value $a_\mathrm{c}$ can be found by calculating the point at which the symmetric solution $\xst=1/2$ changes from being stable to being unstable, i.e., by solving the equation $\left.\frac{d\GV(x)}{dx}\right|_{x=1/2}=0$. A simple calculation shows that this condition is equivalent to 
\begin{equation}\label{eq:ac}
(1-a_\mathrm{c})\left.\frac{d\PhiV(x)}{dx}\right|_{x=1/2}=2a_\mathrm{c}. 
\end{equation}

\begin{figure}[h!]
\includegraphics[width=0.5\textwidth]{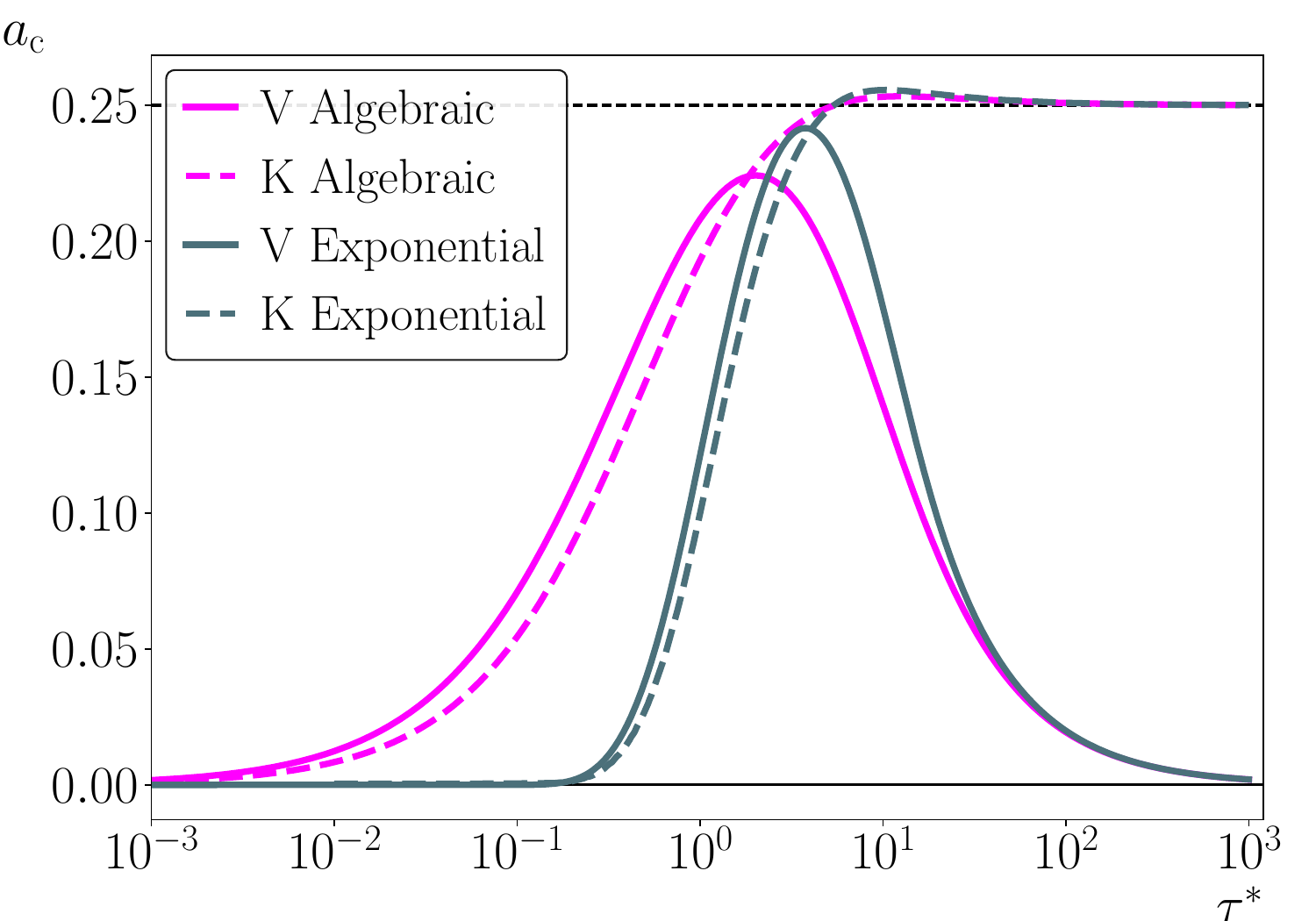}
 \caption{Theoretical curves of the critical noise value $a_\mathrm{c}$ versus the $\tau^*$ parameter in the algebraic and exponential cases, for the noisy voter (V) and kinetic-exchange (K) models. The horizontal lines correspond to the critical value in the aging-less case in the noisy voter (solid line) and in the kinetic-exchange (short-dashed line) models. 
 }\label{fig:VK_acvsts}
\end{figure}

Note that $a_\mathrm{c}$ also appears in the function $\PhiV(x)$. The dependence of the critical value $a_\mathrm{c}$ on the parameter $\tau^*$, obtained by solving numerically the previous equation, is shown in Fig.~\ref{fig:VK_acvsts}. The larger the value of $a_\mathrm{c}$, the larger the region in parameter space in which a consensus-like phase is present. Remarkably, both for the algebraic and the exponential cases, the critical value tends to zero for $\tau^*\to 0$ and $\tau^*\to\infty$, indicating that in those limits the only asymptotic solution is the disordered one. The limit $\tau^*\to\infty$ leads to $q_\tau=1,\,\forall \tau$, which corresponds to the standard version (without aging) of the noisy-voter model. It is known that this model displays a finite-size noise-induced phase transition between disorder and consensus at a critical value of the parameter $a_\mathrm{c}=2/N$. In the thermodynamic limit, it is $a_\mathrm{c}\to0$, in accordance with our result. The reason of the disappearance of the consensus in the limit $\tau^*\to0$ stems from the fact that the social mechanism is activated with a probability $q(\tau)$ that tends to $0$ for all $\tau$, except $\tau=0$, for which $q(0)=q_0>0$. Hence, the only effective updating mechanism acting at all times is random updates, that necessarily lead to disorder. In between these two limits there is an optimal value $\tau^*_\mathrm{c}$ for which $a_\mathrm{c}$ is maximum. For the algebraic case, it is $\tau^*_\mathrm{c} = 2.01$ and the corresponding value of the noise intensity is $a_\mathrm{c} = 0.224$, while for the exponential case we find $\tau^*_\mathrm{c} = 3.77$ and $a_\mathrm{c} = 0.242$. The critical lines $a_\mathrm{c}(\tau^*)$ of the algebraic and exponential cases cross at the point $\tau^*_0 = 2.33$, such that for $\tau^*<\tau^*_0$, the algebraic aging shows a larger consensus region (larger value of $a_\mathrm{c}$) and the opposite for $\tau^*>\tau^*_0$.

\section{Comparison with the noisy kinetic-exchange model with aging}\label{sec:kinetic}

We now compare the main similarities and differences between the noisy voter model and the kinetic-exchange both under the presence of aging.

The kinetic exchange model introduces a third, neutral, value of the state variable, $s_i \in \{-1,0,+1\}$. At variance with the noisy voter model, a macroscopic characterization uses both densities $x^\pm(t)$ of agents in states $\pm1$ at time $t$. By normalization, the density of agents in state $0$ is given by $x^0(t)=1-x^+(t)-x^-(t)$.

Along the lines used for the voter model, and developed in more detail in Ref.~\cite{OurArXiv2023}, it is possible to write down a closed system of equations for the densities $x^\pm(t)$, namely,
\begin{equation}\label{eq:dxdtKP}
\begin{split}
\frac{dx^+}{dt}&=\GK(x^+,x^-),\\ 
\frac{dx^-}{dt}&=\GK(x^-,x^+),\\
\GK(z,w)&=(1-a)\left[z(1-z-w)\PhiK(z+w)-z\, w\,\PhiK(w)\right]+\frac{a}{3}\left(1-3z\right).
\end{split}
\end{equation} 
where 
\begin{equation} \label{eq:PhiK}
\PhiK(x) \equiv 
\frac{\sum_{\tau=0}^\infty q_\tau \FK_\tau(x)}{\sum_{\tau=0}^\infty \FK_\tau(x)}.
\end{equation}
\begin{equation} \label{eq:FKP}
\FK_0(x)=1,\quad \FK_\tau(x)= \prod_{k=0}^{\tau-1} \gamma_3(q_k\, x,a),\quad \tau\ge 1,
\end{equation}
where $\gamma_n(z,a)$ is given by Eq.~(\ref{eq:Xi_def}). Although the dynamical system for the kinetic model, Eqs.~\eqref{eq:dxdtKP}, is very different from that of the voter model, Eqs.~\eqref{eq:dxdtVP}, a very similar structure is found for the functions $\PhiV(x)$ and $\PhiK(x)$. As explained in Appendix~\ref{sec:app:Ftau}, both functions are expressible in terms of hypergeometric functions for the algebraic dependence of the aging probabilities and can be computed using a very efficient numerical algorithm in the case of the exponential dependence. 

Let us summarize now the main results of the analysis of the dynamical system Eqs.~\eqref{eq:dxdtKP}. In the aging-less case, $\PhiK(x)=q_0$, the dynamical equations always admit the steady-state solution $\xst^+=\xst^-=1/3$. This solution becomes unstable for values of the noise intensity $a$ less than the critical value $a_\mathrm{c}=\frac{q_0}{3+q_0}$, where a pair of non-symmetric solutions $\xst^+\ne \xst^-$ emerge. The behavior of the magnetization $m_\mathrm{st}=|\xst^+-\xst^-|$ is given by
\begin{equation}
m_\mathrm{st}= \begin{cases}
{
 \displaystyle
 \frac{\sqrt{(q_0-(3+q_0)a)(3q_0-(1+3q_0)a)}}{\sqrt{3}\,q_0\,(1-a)}
}, &a\le a_\text{c} \\0, &a\ge a_\text{c}
 \end{cases}
 ,
\label{eq:noaging-mag}
\end{equation}
a result already obtained by Crokidakis~\cite{Crokidakis2014} in the case $q_0=1$.

When aging is included, the basic structure of the solution remains~\cite{OurArXiv2023}: there is always a symmetric phase $\xst^+=\xst^-$ steady-state solution, which is stable for $a~>~a_\mathrm{c}(\tau^*)$. For $a~<~a_\mathrm{c}(\tau^*)$ this solution becomes unstable and a pair of non-symmetric stable solutions emerge. It is not possible to give explicit expressions for the solutions $\xst^\pm$ for an arbitrary function $\PhiK(x)$ and they have to be determined numerically. The stability of these solutions is determined also numerically from the sign of the eigenvalues of the Jacobian matrix
\begin{equation}
\left.
\begin{pmatrix}
 \dfrac{\partial\dxst^+}{\partial x^+}&\dfrac{\partial\dxst^-}{\partial x^+}\\[15pt]
 \dfrac{\partial\dxst^+}{\partial x^-}& \dfrac{\partial\dxst^-}{\partial x^-}
 \end{pmatrix}
 \right|_{x^\pm=\xst^\pm}
\end{equation}
evaluated at the fixed points. 

\begin{figure}[b!]
 \includegraphics[width=0.45\textwidth]{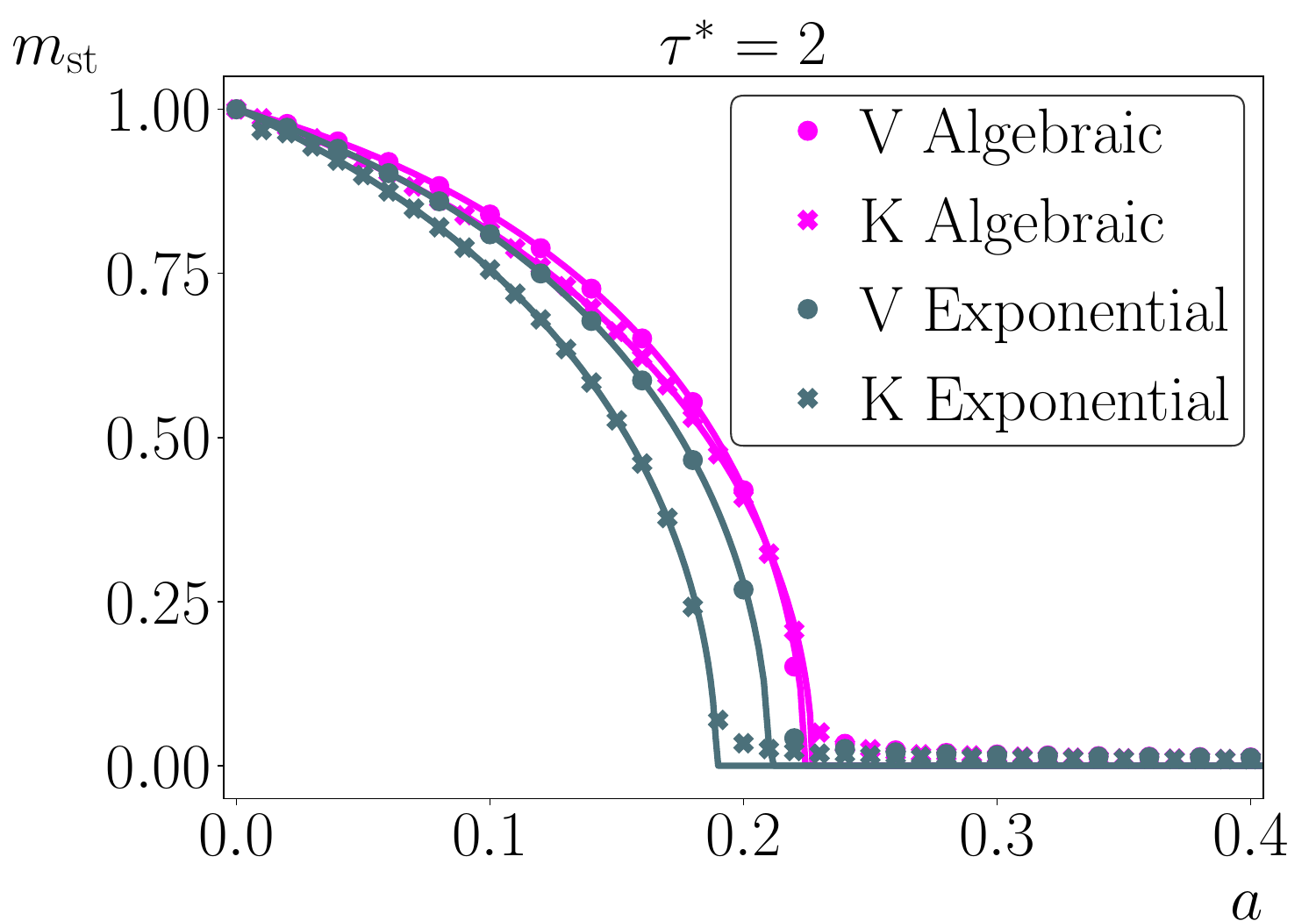}
 \includegraphics[width=0.45\textwidth]{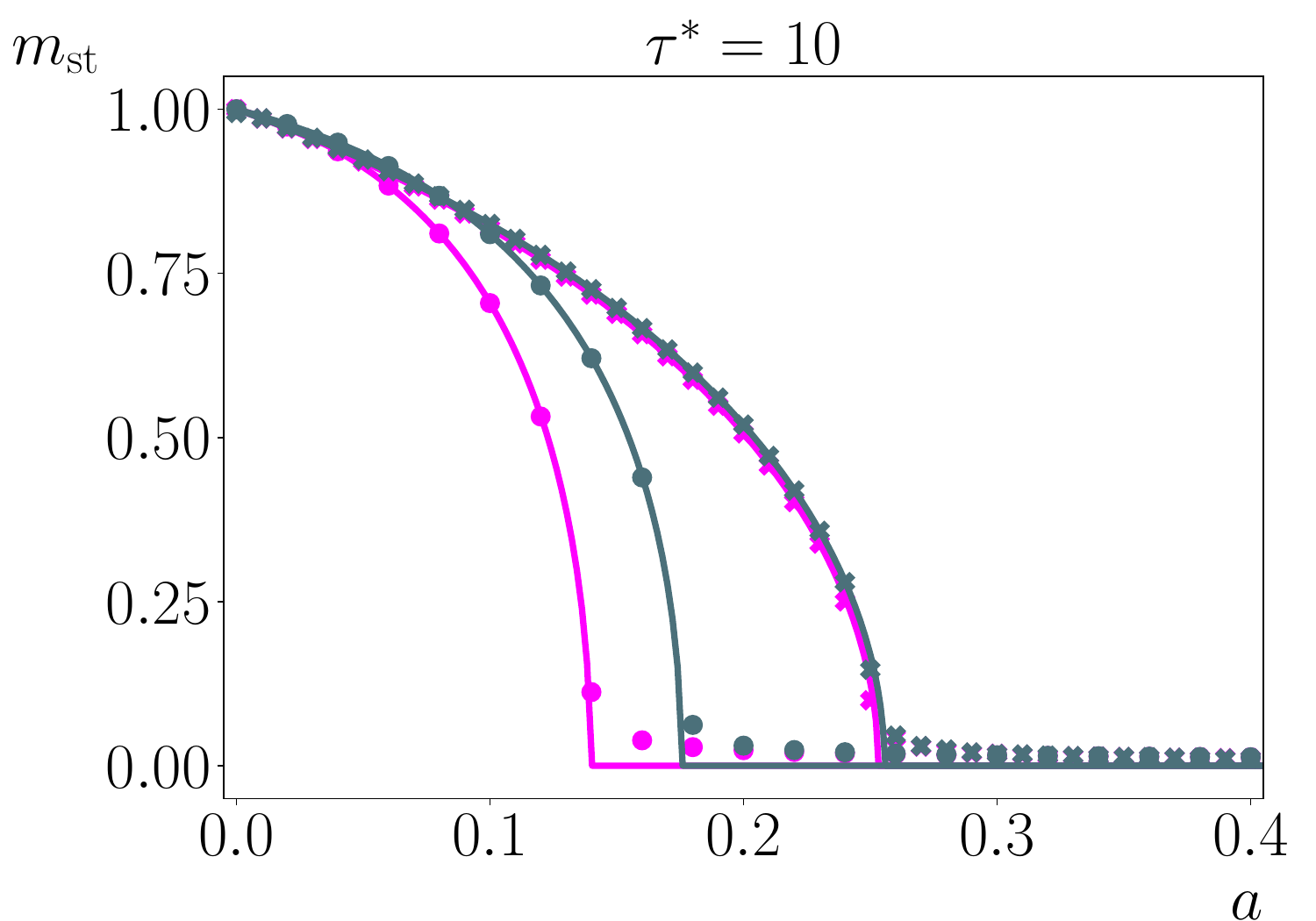}
\caption{Comparison of the noisy voter model (V) and the noisy kinetic-exchange model (K), both under the influence of aging. Magnetization $m_\mathrm{st}$ as a function of the noise intensity $a$ in the algebraic and in the exponential cases for two values of the $\tau^*$ parameter: $\tau^* = 2$ (left panel) and $\tau^* = 10$ (right panel). Results of the mean-field theory are plotted with solid lines, while those of numerical simulations are displayed with symbols. The simulations have been performed with $10^4$ agents and averaged over $5\times 10^6$ MCS after $5\times 10^6$ thermalization MCS.}
\label{fig:VK_mstvsa}
\end{figure}

In Fig.~\ref{fig:VK_mstvsa} we show the results of the magnetization $m_\mathrm{st}$ as a function of the noise intensity $a$ for two values of the $\tau^*$ parameter for the noisy voter ($m_\mathrm{st} = |2 x_\mathrm{st} - 1|$) and for the kinetic-exchange ($m_\mathrm{st} = |x_\mathrm{st}^+-x_\mathrm{st}^-|$) models with aging. In both cases we find excellent agreement between theoretical results and numerical simulations using the stochastic rules of the processes.

As can be seen in Fig.~\ref{fig:VK_acvsts}, for sufficiently small values of $\tau^*$ (strong or moderate aging, e.g., $\tau^*=2$ in Fig.~\ref{fig:VK_mstvsa}), the most determining factor for the existence of a given consensus region is not the type of model considered (either voter or kinetic-exchange), but the type of aging kernel. In particular, algebraic aging, which decays most slowly with $\tau$, is the one that most enhances this effect. On the other hand, for large values of $\tau^*$ (weak aging, e.g., $\tau^*=10$ in Fig.~\ref{fig:VK_mstvsa}), the kinetic-exchange model gives rise to considerably larger consensus regions than the voter model, regardless of the type of aging considered (be algebraic or exponential), consistently with the fact that in the aging-less limit ($\tau^*\to \infty$) the kinetic model has a non-null critical point in the thermodynamic limit, in contrast to the voter model.

The most interesting result for the kinetic model is the non-monotonic behavior of the magnetization: the critical value $a_\mathrm{c}(\tau^*)$ exceeds the value corresponding to the aging-less case $a_\mathrm{c}=1/4$ and it then approaches asymptotically to $1/4$. As can be seen in Fig.~\ref{fig:VK_acvsts}, this phenomenon occurs for both profiles of aging. Moreover, as in the noisy voter model with aging, also in the kinetic model there is a value $\tau_0^*$ such that for $\tau^* < \tau_0^*$ the algebraic aging gives rise to a larger consensus region than the exponential aging, while for $\tau^* > \tau_0^*$ the situation is reversed. For the noisy kinetic-exchange model it is $\tau_0^* = 5.40$.

\section{Final remarks}\label{sec:final}
We have considered two models of opinion formation, namely, the noisy versions of the 2-state voter and 3-state kinetic-exchange models, and introduced the effect of aging through the probability of change $q_\tau$ acting on the interaction between agents. Moreover, we considered two different families of kernels for $q_\tau$: either algebraically or exponentially decaying with $\tau$. 

A mean-field description has been given, for generic form of $q(\tau)$, unifying previous results for the two models~\cite{Artime2018,OurArXiv2023} and extending part of them to embrace a wider family of models with $n$ opinion states. This description relies on obtaining the evolution equations for the density of agents in each opinion state $s$ and with given age $\tau$. Then, the steady state solutions of this system of (infinite) equations and the analysis of its stability (via an adiabatic approximation valid when separation of timescales holds) allowed to obtain the plots of the order parameter $m_{st}$ vs. the noise probability $a$, in agreement with agent-based simulations in a complete graph.

We have considered particularly the algebraic form $q_\tau = (q_\infty \tau+q_0 \tau^*)/(\tau+\tau^*)$, and the exponential form $q_\tau = q_\infty + (q_0-q_\infty)\exp(-\tau/\tau^*)$, with $0\le q_\infty < q_0 \le 1$. In numerical examples, we focused on the case $(q_0,q_\infty)=(1,0)$. Our results indicate that qualitatively similar tendencies emerge for both kernels in both models as follows. For very large $\tau^*$ (weak aging), there is a good agreement with the aging-less situation, as expected. For all values of $\tau^*$, the magnetization $m_\mathrm{st}$ as a function of the noise intensity $a$ displays a continuous transition from order to disorder at a critical value $a_\text{c}$. 
In all the analyzed cases we observe an optimal value of $\tau^*$ to obtain consensus (maximizing this critical value $a_\mathrm{c}$). It is also noteworthy that, while both models produce different results under weak aging, when aging is strong or moderate, the critical value $a_c$ becomes nearly insensitive to the particular social interaction rule $(s_i,s_j)$ and number of opinion states $n$.
 
An interesting continuation would be to go beyond all-to-all interactions and to study aging effects on the opinion dynamics in random networks. Moreover, other forms of the aging kernel might also bring new features. 

\acknowledgments{
Partial financial support has been received from the Agencia Estatal de Investigaci\'on (AEI, MCI, Spain) \\MCIN/AEI/10.13039/501100011033 and Fondo Europeo de Desarrollo Regional (FEDER, UE) under Project APASOS (PID2021-122256NB-C21) and the María de Maeztu Program for units of Excellence in R\&D, grant CEX2021-001164-M. 
C.A. also acknowledges partial support received from
 Conselho Nacional de Desenvolvimento Científico e Tecnológico (CNPq)-Brazil (311435/2020-3) and Fundação de Amparo à Pesquisa do
Estado de Rio de Janeiro (FAPERJ)-Brazil (CNE E-26/201.109/2021). 
}
\bibliography{Aging_total.bib}

\clearpage

\appendix
\section{Rate equations}
\label{app:rates}
In the general case of $n$ possible state values $s$, the rate equations for the density $x^s_\tau$ of agents in state $s$ and age $\tau$ are
\begin{equation} \label{eq:app:dxdt_tau}
 \frac{dx_{\tau}^{s}}{dt}=\Omega_{ss}(\tau-1)-\sum_{s'}{\Omega_{ss'}(\tau)},\quad \tau\ge 1,
\end{equation}
where $\Omega_{ss'}(\tau)$ is the transition rate from state $s$ to state $s'$. The first term accounts for those updates that did not result in a change of state and hence increase their internal time from $\tau-1$ to $\tau$, while the second accounts for those that resulted in a change of state and therefore resulted in a change $\tau\to 0$ of the internal time. For $\tau=0$ the rate equation includes all those changes that result in a reset of the internal time,
\begin{equation} \label{eq:app:dxdt_0}
 \frac{dx_{0}^{s}}{dt}=\sum_{\tau=0}^{\infty}\sum_{s'\neq s}{\Omega_{s's}(\tau)}-\sum_{s'}{\Omega_{ss'}(0)}.
\end{equation}
Summing Eqs.~(\ref{eq:app:dxdt_tau}, \ref{eq:app:dxdt_0}) over all values of $\tau$, one obtains the rate equations for the density of agents in each state $s$ as $x^s=\sum_{\tau=0}^{\infty}x^s_\tau$. We derive now the rate expressions for the noisy voter. A similar treatment for the kinetic-exchange model can be found in Ref.~\cite{OurArXiv2023}.

\subsection[]{Noisy voter model with aging} 

\label{sec:app:nvm}
The state can take two possible values $s\in\{-1,+1\}$. For the all-to-all connectivity adopted in this paper the set of transition rates $\Omega_{ss'}$ are
\begin{equation}\label{eq:Vomegas}
\begin{split}
\Omega_{--}(\tau)&= x_{\tau}^{-}\left(\frac{a}{2} +(1-a)(1-x^+ q_\tau) \right), \\
\Omega_{-+}(\tau)&= x_{\tau}^{-}\left(\frac{a}{2} +(1-a)x^+ q_\tau \right),\\
\Omega_{+-}(\tau)&= x_{\tau}^{+}\left(\frac{a}{2} +(1-a)x^-q_\tau \right),\\
\Omega_{++}(\tau) &= x_{\tau}^{+}\left(\frac{a}{2} +(1-a)(1-x^-q_\tau)\right).
\end{split}
\end{equation}
In the main text we have used the notation $x$ for $x^+$ and $1-x$ for $x^-$. We show now the derivation of the expression for the first of these rates $\Omega_{--}(\tau)$: The probability that an agent with internal time $\tau$ and state value $s=-1$ remains $-1$ requires first to select an agent in that state (probability $x^-_\tau$), then if the social rule is chosen (probability $1-a$), the copying mechanism activates with probability $q_\tau$ and the selected neighbor must be in state $-1$ (probability $x^-$) but, if the copying mechanism is not activated (probability $1-q_\tau$) the state remains $-1$. If, on the other hand, the idiosyncratic rule is activated (probability $a$), and the new state is chosen equal to $-1$ with probability $1/2$. This leads to
\begin{equation}
\Omega_{--}(\tau)=x^-_\tau \left((1-a) ( x^- q_\tau +(1-q_\tau))+ a\,\frac{1}{2} \right),
\end{equation}
which, after replacement of $x^-=1-x^+$ leads to the first of Eqs.~\eqref{eq:Vomegas}. Other transition rates in these equations are obtained similarly.

From these rates and Eqs.~(\ref{eq:app:dxdt_tau}, \ref{eq:app:dxdt_0}) one can derive the rate equations for $x_{\tau}^s$ as
\begin{equation} \label{eq:Vratesi}
 \begin{split}
 \frac{dx_\tau^-}{dt} &=\Omega_{--}(\tau-1)-x_\tau^-,\\
 \frac{dx_\tau^+}{dt} &=\Omega_{++}(\tau-1)-x_\tau^+,
 \end{split}
\end{equation}
for $\tau\geq1$, and
\begin{equation} \label{eq:Vratesi0}
 \begin{split}
 \frac{d x^{-}_0}{dt} &
 = \frac{a}{2} x^+ +(1-a)x^-y^+ -x_0^-,
\\
\frac{d x^{+}_0}{dt} &
 = \frac{a}{2} x^- +(1-a)x^+y^- -x_0^+,
 \end{split}
\end{equation}
for $\tau=0$, where we have defined 
\begin {equation} \label{eq:app:ys}
 y^s \equiv \sum_{\tau=0}^{\infty} q_\tau x^s_\tau.
\end{equation}
To obtain a closed equation for the time evolution of the densities $x^s$ we use an adiabatic approximation whereby we assume that the time derivative Eq.~\eqref{eq:Vratesi} of the age-dependent density $x^s_\tau$ can be set to zero, allowing to express $x^\pm_\tau$ in terms of $x^\pm_{\tau-1}$. The solution of this recursive relation is
\begin{equation} \label{eq:xtau}
\begin{split}
x_\tau^-&=x_0^- F_\tau(x^+),\\
x_\tau^+&=x_0^+ F_\tau(x^-),
\end{split}
\end{equation}
with
\begin{equation} \label{eq:Ftau}
F_0(x)=1, \hspace{1cm}
F_\tau(x)\equiv \prod_{k=0}^{\tau-1}\gamma_2(q_k\, x,a),\quad \tau\ge 1,
\end{equation}
and the function $\gamma_2$ is defined in Eq.~\eqref{eq:Xi_def}.
Summing Eq.~(\ref{eq:xtau}) over all $\tau\ge 0$, we obtain
\begin{equation} \label{eq:xxx}
\begin{split}
x^- &= x_0^-\sum_{\tau=0}^{\infty}F_\tau(x^+),\\ 
x^+ &= x_0^+\sum_{\tau=0}^{\infty}F_\tau(x^-),
\end{split}
\end{equation}
which substituted in Eqs.~\eqref{eq:xtau} leads to

\begin{equation} \label{eq:app:Vxtau}
\begin{split}
x_\tau^-&=x^- \frac{F_\tau(x^+)}{\sum_\tau F_\tau(x^+)},\\
x_\tau^+&=x^+ \frac{F_\tau(x^-)}{\sum_\tau F_\tau(x^-)},
\end{split}
\end{equation}
which are now expressed in terms of the global variables $x^\pm$.

Adding Eqs.~(\ref{eq:Vratesi}) and (\ref{eq:Vratesi0}) over all values of $\tau$, we obtain the corresponding equations for the density $x^s$ of each state $s$. 
\begin{equation} \label{eq:app:Vrates+-}
 \begin{split}
\frac{d x^{-}}{dt} &=
\frac{a}{2}( x^+ -x^-) 
+ (1-a)( x^-y^+ -x^+ y^-) , \\
\frac{d x^{+}}{dt} &=
\frac{a}{2}( x^- -x^+) 
+ (1-a)(x^+ y^- - x^- y^+).
\end{split}
\end{equation}
Moreover note that one of the two equations can be eliminated since $x^+ +x^- =1$. 
Then, in order to obtain a closed evolution equation for the global variable $x^+$, we need to express the variables $y^s$ appearing in Eqs.~\eqref{eq:app:Vrates+-} in terms of $x^+$. 
This can be done by using Eq.~(\ref{eq:app:Vxtau}) into Eqs.~\eqref{eq:app:ys}, yielding
\begin{equation} \label{eq:Vyyy}
y^- =(1-x^+) \Phi(x^+), \;\;\;\;\; y^+ =x^+ \Phi(1-x^+), 
\end{equation}
where we have introduced the function 
\begin{equation} \label{app:eq:Phis}
\Phi(x) \equiv 
\frac{\sum_{\tau=0}^\infty q_\tau F_\tau(x)}{\sum_{\tau=0}^\infty F_\tau(x)} .
\end{equation}
Let us note here, for consistency, that in the aging-less case, $ q_\tau=1$, it is $ \Phi(x)= 1$ and, hence, $y^s=x^s$, for $s=-1,+1$. 

Replacement of Eqs.~\eqref{eq:Vyyy} in Eqs.~\eqref{eq:app:Vrates+-} leads to the rate equation for $x\equiv x^+$, Eq.~\eqref{eq:dxdtVP}. Note that the function $\Phi(x)$ depends on the aging profile $q_\tau$ both through the explicit dependence of Eq.~\eqref{app:eq:Phis} and in the expression of $F_\tau(x)$ of Eq.~\eqref{eq:Ftau}. As shown in Appendix~\ref{sec:app:Ftau} the function $\Phi(x)$ can be related to hypergeometric functions in the case of an algebraic dependence of the aging probability. 

\section{Calculation of sums involving $F_\tau(x)$}
\label{sec:app:Ftau}

\subsection[]{Algebraic aging} \label{sec:app:Ftau_alg}
In the case of a rational function of the age of the general form 
\begin{equation}
 q_{\tau}=\frac{q_{\infty}\tau+q_0\tau^*}{\tau+\tau^*},
\end{equation}
where $\tau^*>0$ and $0\leq q_{\infty} < q_0$, the function $F_\tau(x)$ defined is given by
\begin{equation} \label{eq:Ftau_def}
F_\tau(x)\equiv \prod_{k=0}^{\tau-1}\gamma_n(q_k\, x,a)=\gamma_n(q_{\infty} x,a)^{\tau }\frac{ \left(\tau^*\xi_n(x,a)\right)_{\tau }}{(\tau^*)_{\tau }},\quad \tau\ge 1,
\end{equation}
where
\begin{equation}
 \xi_n(x,a)\equiv\frac{\gamma_n(q_0 x,a)}{\gamma_n(q_\infty x,a)},
\end{equation}
with the function $\gamma_n(z,a)$ defined in Eq.~\eqref{eq:Xi_def}, and
 $(z)_\tau\equiv \Gamma(z+\tau)/\Gamma(z)$ is the Pochhammer symbol.

In order to compute the function $\Phi(x)=\displaystyle\frac{\sum_{\tau=0}^\infty q_\tau F_\tau(x)}{\sum_{\tau=0}^\infty F_\tau(x)}$, we need the following sums: 
\begin{align} \label{eq:NKM_total_sumF}
 \sum_{\tau=0}^{\infty}F_{\tau}(x)=&\,_2F_1\left(1,\tau^*\xi_n(x,a);\tau^*;\gamma_n(q_\infty x, a) \right),\\
 \sum_{\tau=0}^{\infty}q_{\tau}F_{\tau}(x)=&q_0 \,_2F_1\left(1,\tau^*\xi_n(x,a);1+\tau^*;\gamma_n(q_\infty x, a) \right)+ \nonumber\\
 &\frac{q_{\infty} }{1+\tau^*}\gamma_n(q_0 x,a)\, _2F_1\left(2,1+\tau^*\xi_n(x,a);2+\tau^*;\gamma_n(q_\infty x, a)\right). \label{eq:NKM_total_sumqF}
\end{align}
In this work we cover the algebraic case corresponding to $q_0=1$ and $q_\infty=0$.

\subsection[]{Exponential aging} \label{sec:app:Ftau_exp}
    In the definition of $F_\tau(x)$, Eq.~(\ref{eq:Ftau_def}), where $\gamma_n(z,a)$ is given by Eq.~(\ref{eq:Xi_def}), the substitution of $q_\tau=e^{-\tau/\tau^*}$ yields 
\begin{equation}
F_\tau(x)=\biggl(\underbrace{1-\frac{n-1}{n}a}_\alpha\biggr)^\tau
\biggl( 
\underbrace{n x\frac{1-a}{n-(n-1)a}}_b;e^{-1/\tau^*}
\biggr)_\tau,
\end{equation}
where 
$(r;s)_0=1$ and $(r;s)_k=\prod_{i=0}^{k-1}(1-rs^i)$
for $k\ge 1$, is the $q$-Pochhammer symbol, 
and where $\alpha \in(1/n,1)$ and $b\in(0,1)$. 
Therefore, the function $\Phi(x)$ is given by 
\begin{equation}
\label{eq:Phi_sum}
\Phi(x)=\frac{\sum_{\tau=0}^\infty q_\tau F_\tau(x)}{\sum_{\tau=0}^\infty F_\tau(x)}.
\end{equation}
A way to compute numerically the infinite series in the numerator and denominator of Eq.~(\ref{eq:Phi_sum}) is to cut-off them by finite sums up to an upper index $\tau=M$. However, the convergence seems to be slow and large values of $M$ are needed for a precise calculation, specially for small values of $\tau^*$ or $a$. To overcome this difficulty, an efficient procedure has been described in Ref.~\cite{OurArXiv2023}. It starts by introducing the function 
\begin{eqnarray}\label{eq:phidef}
\phi(z,b,s)=\sum_{\tau=0}^\infty (b;s)_\tau\,z ^\tau\,, 
\end{eqnarray}
from where Eq.~(\ref{eq:Phi_sum}) reads
\begin{equation}
\label{eq:Phi_sum2}
\Phi(x)=\frac{\phi(\alpha s,b,s)}{\phi(\alpha,b,s)}, \quad s=e^{-1/\tau^*}\in(0,1).
\end{equation}
Then, we use the iteration relation that follows from its definition, Eq.~\eqref{eq:phidef}, namely,

\begin{equation}\label{recursion}
\phi(\alpha,bs^{k},s)=1+\alpha(1-bs^{k})\,\phi(\alpha,bs^{k+1},s), \quad k=L,L-1,\dots,0
\end{equation}
with the initial condition \begin{equation}\phi(\alpha,bs^{L+1},s)=\frac{1}{1-\alpha},
\end{equation}
that results from the approximation $bs^{L+1}\approx 0$ and the use of $(0;s)_k=1$ in the definition~\eqref{eq:phidef}.
We have taken $L$ such that $bs^{L+1}<\epsilon=10^{-12}$, or $L\sim\log(\epsilon/b)/ \log{s}=\tau^*\log(b/\epsilon)$ although other smaller values of $\epsilon$ produced the same results.

\end{document}